\documentclass[%
 reprint,
 amsmath,amssymb,
 aps,
prr,
]{revtex4-2}

\usepackage{graphicx}% Include figure files
\usepackage{dcolumn}% Align table columns on decimal point
\usepackage{bm}% bold math
\usepackage{subfig}
\usepackage{algorithm2e}
\usepackage{xcolor}
\newcommand{\rev}[1]{\textcolor{black}{#1}}%red
\newcommand{\revb}[1]{\textcolor{black}{#1}}%blue

\begin{document}

\preprint{APS/123-QED}

\title{Hyperbolic embedding of brain networks as a tool for epileptic seizures forecasting}% Force line breaks with \\

\author{Martin Guillemaud}
%\email{martin.guillemaud@gmail.com}
\altaffiliation{These authors contributed equally to this work}
 \affiliation{Paris Brain Institute (ICM), CNRS, Inserm, Sorbonne University, Inria-Paris. Pitié Salpêtrière University Hospital. Paris, France }%Lines break automatically or can be forced with \\
\author{Louis Cousyn}%
 \altaffiliation{These authors contributed equally to this work}
\affiliation{%
Paris Brain Institute (ICM), Sorbonne Université, CNRS UMR, Inserm.\\ AP-HP, Department of Neurology, Epilepsy Unit,
Center of Reference for Rare Epilepsies, ERN EPICARE, Pitié Salpêtrière University Hospital, Paris, France
}%
\author{Vincent Navarro}%
\affiliation{%
Paris Brain Institute (ICM), Sorbonne Université, CNRS UMR, Inserm.\\ AP-HP, Department of Neurology, Epilepsy Unit,
Center of Reference for Rare Epilepsies, ERN EPICARE, Pitié Salpêtrière University Hospital, Paris, France
}%
\author{Mario Chavez}
\affiliation{CNRS, Pitié Salpêtrière University Hospital. Paris, France \\}

%\affiliation{
%\affiliation{
%\affiliation{%


\date{\today}% It is always \today, today,

\begin{abstract}
The evidence indicates that intracranial EEG connectivity, as estimated from daily resting state recordings from epileptic patients, may be capable of identifying preictal states. In this study, we employed hyperbolic embedding of brain networks to capture non-trivial patterns that discriminate between connectivity networks from days with (preictal) and without (interictal) seizure. A statistical model was constructed by combining hyperbolic geometry and machine learning tools, which allowed for the estimation of the probability of an upcoming seizure. The results demonstrated that representing brain networks in a hyperbolic space enabled an accurate discrimination (85\%) between interictal (no-seizure) and preictal (seizure within the next 24 hours) states. The proposed method also demonstrated excellent prediction performances, with an overall accuracy of 87\% and an F1-score of 89\% (mean Brier score and Brier skill score of $0.12$ and $0.37$, respectively). In conclusion, our findings indicate that representations of brain connectivity in a latent geometry space can reveal a daily and reliable signature of the upcoming seizure(s), thus providing a promising biomarker for seizure forecasting.
\end{abstract}

%\keywords{Suggested keywords}%Use showkeys class option if keyword display desired
\maketitle
%\tableofcontents
\section{Introduction}

In the last decades, complex networks have provided an increasingly challenging framework for the study of connected systems (from social sciences to biology and physics), based on the interplay between the wiring architecture and the dynamical properties of the coupled elements~\cite{11_BOCCALETTI2006175}. In neurosciences, the representation of brain networks as graphs allows to better describe their non-trivial connectivity properties in a compact and objective way~\cite{12_REIJNEVELD20072317,13_Bullmore2009}. In this mapping of brain data (e.g. scalp or intracranial electroencephalography or magnetoencephalography) to networks, nodes usually represent brain regions or recording sites (e.g., electrodes, sensors), and edges or links indicate functional connections between them, based on an estimated statistical relationship between the recorded signals. Regardless of the modality of acquisition, the use of graph analysis in neurosciences has become essential to characterize pathological or physiological states in terms of connectivity brain networks~\cite{12_REIJNEVELD20072317, 13_Bullmore2009, 14_DeVicoFallani2014}. The application of different graph metrics, including node's degree, centrality, communicability, and efficiency, has yielded insights into brain function in both healthy and pathological conditions~\cite{14_DeVicoFallani2014}. 

Epilepsy is a neurological disorder nowadays conceptualized as a network disease with functionally and/or structurally aberrant connections on virtually all spatial scales~\cite{16_Kramer2012}. According to this concept, a large-scale epileptic network comprises brain areas that can generate and sustain normal and physiological dynamics during the seizure-free interval, and are mainly involved in the generation, maintenance, spread, and termination of pathophysiological activities such as seizures~\cite{16_Kramer2012, 17_Diessen2013}. Network connectivity analysis in epilepsy has provided valuable information on seizure onset and propagation, as well as on the functional organization of the brain during the seizure-free interval~\cite{17_Diessen2013}. During seizures, brain networks have been found to display a more regular structure with less variability~\cite{16_Kramer2012, 17_Diessen2013}. Nevertheless, current representations of brain networks fail to provide reliable biomarkers to predict seizure onset or to estimate a risk of seizure occurrence.~\cite{15_Kuhlmann2018, 16_Kramer2012, 17_Diessen2013}.  

Embedding (or vector space) methods identify a lower-dimensional space in which high-dimensional complex data can be represented. Modern dimensionality reduction methods learn similarities and proximities between points distributed over a hidden manifold in a multidimensional feature space. These methods then preserve, embed, and visualize the data in a low-dimensional space. Although Euclidean geometry serves as a standard framework for studying our physical reality, an increasing body of evidence suggests that non-Euclidean geometries are a more appropriate framework for capturing non-trivial features (e.g., hierarchical or multiscale structure) often observed in different biological networks, such as the olfactory space, cell development data, single-cell sequencing data, and the brain connectivity \cite{19_Boguna2021}. 

The prevailing network embedding methodologies assume a zero-curvature (or flat) space and evaluate the distance between embedded nodes according to Euclidean metrics. These approaches, however, are limited when dealing with complex hierarchical structures, as the resulting pairwise distances in the embedded spaces are substantially distorted~\cite{19_Boguna2021}. In network analysis, the question of whether it exists a hidden (latent) non-Euclidean geometry from which complex connectivity emerges has recently attracted the attention of the scientific community and is getting more and more ground~\cite{21_Allard2020, 22_Zheng2020}. Recently, spaces with negative curvature (hyperbolic geometries) have attracted a lot of attention as they enable low-distortion embeddings of hierarchical or multi-scale connectivity structures~\cite{19_Boguna2021}.  The possibility to find an effective congruency between brain network characteristics and its representation in hyperbolic spaces offers thus the possibility to understand its structure and address the study of brain organization in a novel and promising framework. In some cases, endowed metrics of such spaces allow the use of machine learning tools to perform statistical analysis of the embedded points (e.g. clustering, prediction, \ldots)~\cite{20_Hofmann2008}. 

\rev{Current methods for predicting epileptic seizures largely rely on long-term electroencephalographic (EEG) recordings, obtained either from scalp or intracranial sources. These techniques impose significant demands on both data collection and processing~\cite{Ilyas2023, Cook2013}. These models are designed to identify specific preictal changes associated with each seizure, without considering fluctuations caused by shifts in vigilance states. As a result, they require either real-time monitoring of vigilance levels or the incorporation of diverse interictal reference states While some methods achieve high sensitivity, they are often associated with high false positive rates, resulting in frequent false alarms that can diminish their clinical utility~\cite{Usman2019, Rasheed2021}. Additionally, performance varies significantly across studies due to differences in datasets, feature selection, and model architectures, underscoring the absence of a standardized, generalizable solution for seizure prediction. Despite these challenges, recent advancements in brain network analysis have created new opportunities for developing more efficient and interpretable predictive models.}

\rev{Previous studies~\cite{le2005preictal, van2014functional} have shown that brain connectivity matrices derived from intracranial EEG data can effectively identify preictal states. Building on these findings, we introduce a novel seizure forecasting approach that addresses key limitations of existing methods. In contrast to continuous monitoring systems, our approach is centered on two primary objectives: (1) assessing whether patient-specific EEG connectivity during vigilance-controlled resting-state periods differs between seizure and non-seizure days, and (2) prospectively evaluating the utility of these differences for forecasting daily seizure risk through calibrated predictive models.  To estimate seizure risk, we utilize a hyperbolic network embedding method to represent patients’ functional connectivity into a low-dimensional space. This representation capture subtle changes in brain networks associated with seizure occurrence. Our method demonstrates superior performance compared to various forecasting models, including conventional machine learning techniques. By identifying robust biomarkers for seizure prediction, our approach requires less training data while achieving enhanced accuracy, offering a practical and efficient pathway toward improved epilepsy management.}

%Previous studies~\cite{le2005preictal, van2014functional} showed that brain connectivity matrices, estimated from intracranial electroencephalography (EEG) recordings, allowed the identification of preictal states. In the present study we used network embedding methods to predict the risk of seizure occurrence. Our approach aimed at exploring hyperbolic geometry to represent functional brain networks of patients with epilepsy, and to work in a more adapted space representation to unveil patterns that could potentially result in robust biomarkers for seizure forecasting. 

\section{ Material and method}
\subsection{\label{sec:Data_Acqu}Data acquisition}
The EEG dataset used in this study was previously presented in~\cite{24_Cousyn2023}. Briefly, daily 10-minute resting-state intracranial EEG recordings were obtained in 10 patients (mean age 30.7 years) with drug-resistant focal epilepsy from January 2019 to July 2021 in the Epilepsy Unit of the Pitié-Salpêtrière Hospital (Paris, France). The study was conducted in accordance with the Helsinki Declaration and approved by an institutional review board (project C11-16 and C19-55 of the French National Institute of Health and Medical Research sponsor). \rev{The implanted brain regions and the number of electrodes vary between patients, with a median of 20 electrodes per patient and a range from 10 to 62. The original study associated with this cohort provides detailed demographic and clinical information for each patient~\cite{24_Cousyn2023}.}

The mean number of daily recordings per patient was 11. Each daily recorded period was labeled as ``preictal'' in case an electro-clinical seizure occurred in the next 24 hours, or ``interictal'' otherwise. Connectivity graphs were derived from matrices of phase locking values (PLV) estimated between pairs of EEG signals during 20-second non-overlapping epochs (resulting in 30 connectivity matrices per day). Synchrony (PLV) values were obtained in the typical EEG frequency bands: delta ($\delta$) 1-4~Hz, theta ($\theta$) 4-8~Hz, alpha ($\alpha$) 8-13~Hz, beta ($\beta$) 13-30~Hz, low gamma (low $\gamma$) 30-49~Hz and high gamma (high $\gamma$) 51-90~Hz. Only contacts in the gray matter were considered and a bipolar montage between adjacent contacts was applied.

\subsection{Hyperbolic embedding of brain networks}
In this study, we employed hyperbolic geometrical space for the representation of brain networks. Compared with graph mappings in Euclidean spaces, hyperbolic embeddings exhibit low distortion and can unfold network properties, such as clustering or hierarchical community structure \cite{3_pmlr-v80-nickel18a, 4_Saxena2020GraphCurvature, 5_SHARPEE2019ArgHyp}, which could reveal key structural principles underlying the organization of the brain. The main steps required to embed the brain networks into this geometric space include: i) filtering the connectivity matrix, ii) embedding the network into the geometric space, and iii) aligning the embedding in the hyperbolic space. 

\emph{Networks filtering}. The initial stage of network embedding consists of filtering the PLV matrices by applying a threshold to cancel a percentage of the weakest values. Here, we followed Ref.~\cite{8_DeVicoFallani2017}, and the synchrony matrices were filtered such that the final networks reached a predetermined mean degree, set to 4, \rev{as recommended for small-size networks}. Although a spanning tree filter could be applied to ensure a unique acyclic subgraph that connects all nodes~\cite{stam2014trees}, here we preferred the method of Ref.~\cite{8_DeVicoFallani2017} as it emphasizes the intrinsic global and local properties of the network, while preserving its sparsity. 

\emph{Networks embedding}. Hyperbolic embedding methods are becoming increasingly prevalent in the literature on neural circuits~\cite{1_cacciola2017coalescent, 2_Gao2020, 5_SHARPEE2019ArgHyp, 6_Whi2022}. Hyperbolic geometry studies metric spaces with a constant negative curvature that do not conform to Euclidean geometry. Such embedding methods typically project the nodes of a network onto a hyperboloid, which can then be projected onto a two-dimensional hyperbolic space model, such as the Poincaré disk or the so-called Klein model disk. Methods for mapping a network into the hyperbolic disk essentially belong to two families~\cite{Saxena2020hypembsurvey, poincare_fb}: generative model-based (e.g., Mercator~\cite{GarciaMercator1019}) and machine learning-based. Here, we used the coalescent embedding method~\cite{9_Muscoloni2017} to project the brain networks onto the Poincaré disk $\mathbb{D}$. The method belongs to the second family, and we choose it for its remarkable versatility and especially its computational speed~\cite{Longhena_2014}. 

In a nutshell, the method first adjusts the network's edges' weights, giving more weight to those connections that play a greater role in information transmission. Next, the graph is projected onto a 2-dimensional space using a non-linear dimension reduction method such as Isomap or the Laplacian eigenmaps~\cite{18_vonLuxburg2007}. Here we employed the technique of Laplacian Eigenmaps as it is computationally efficient and it optimally preserves nodes' local neighborhood information~\cite{belkin2003laplacian}. At this stage, all nodes have polar coordinates $(r, \theta)$ in the 2-dimensional space. The nodes undergo an equidistant adjustment in which all radii are set to 1 and the angular coordinates of the nodes are modified while maintaining their angular order within the circle. Finally, each vertex is assigned a radius based on its rank in the order of increasing degree. The radius of a node is thus related to its popularity, while the angular distance between two nodes is related to their topological similarity.   

In the Poincaré disk, the hyperbolic distance $\mathrm{dist}_{hyp}(i,j)$ between each pair of nodes $i$ and $j$, assigned with radii $(r_i, r_j)$ and angles $(\theta_i, \theta_j)$ with coordinates $(r_i,\theta_i)$ and $(r_j,\theta_j)$, is computed according to the hyperbolic law of cosines\cite{Kitsak2020hypdist}:
\begin{multline}   
\label{eq_hyp_distance}
    \cosh \mathrm{dist}_{hyp}(i,j) = \cosh r_i \times \cosh r_j \\
    - \sinh r_i \times \sinh r_j \times \cos(\pi-\vert \pi-\vert \theta_i-\theta_j\vert \vert )
\end{multline}

\emph{Graphs alignment}. Prior to comparing groups of embedded networks, it was necessary to correct or align the embeddings, as a minor perturbation could result in a random angular offset to the nodes' positions in the hyperbolic disk. A minor connectivity perturbation may thus result in two embeddings with the same similarities between nodes but with a different structure regarding the coordinates in the disk.~\cite{gursoy2023alignment}. To address this issue, a small rotation angle between two embeddings was added such that the following hyperbolic similarity score was minimized:
\begin{equation}
        \Gamma_{\mathrm{similarity}} =  \sum_{i=0}^{N} \mathrm{dist}_{hyp} \left( \mathrm{Pos_o} (i), \mathrm{Rot}_{\theta}\left[ \mathrm{Pos_p} (i)\right] \right)
    \label{similarity_error_min}
\end{equation}
where $\mathrm{Pos_o}(i)$ and $\mathrm{Pos_p}(i)$ denote the positions of the node $i$ in the first (or reference) and second hyperbolic disk, and $\mathrm{Rot}_{\theta}\left[\cdot\right]$ indicates a rotation in the Poincare’s disc with an angle $\theta \in \left[- \pi, \pi \right[$. In order to facilitate comparison in subsequent steps, all networks from both groups were aligned using the same reference network. 

\subsection{\label{sec:HypScoreDisk} Hyperbolic score disk calculation}
\begin{figure*}
\includegraphics[width=0.9\textwidth]{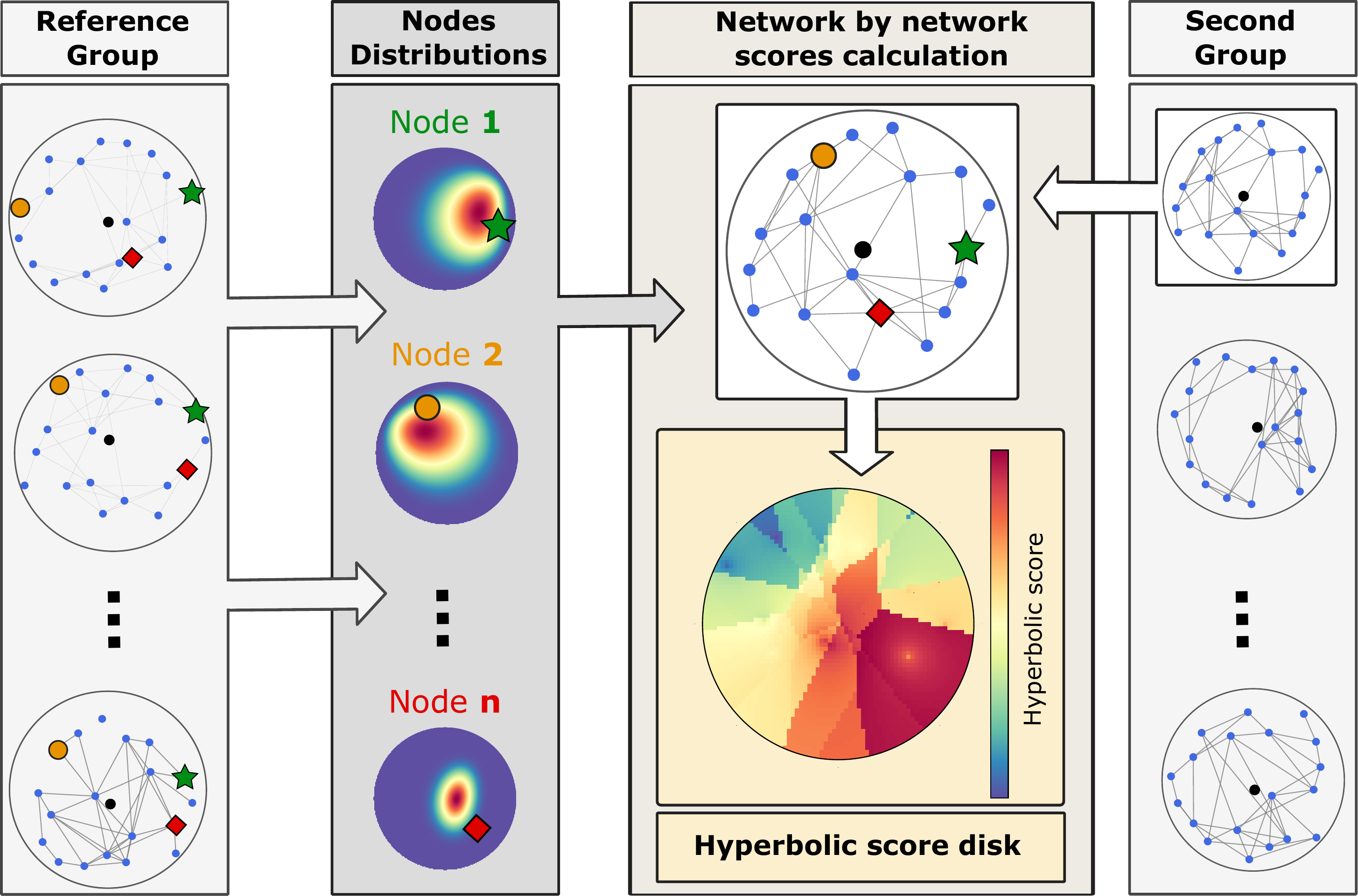}% Here is how to import EPS art
\caption{\label{fig:data_gauss} Calculation of the Hyperbolic Score Disks. First, we computed the hyperbolic Gaussian distributions of the nodes from the reference group (interictal networks). The networks' nodes from the second group are then compared to these distributions, and a score is attributed to each node. Finally, the scores of the nodes are interpolated across the entire hyperbolic disk.}
\vspace{-0.5 cm}
\end{figure*}

Our approach aimed at identifying disparities between preictal and interictal connectivity patterns in the hyperbolic embedding of the brain networks. To achieve this, we first defined a Gaussian model in the hyperbolic space for each embedded node of the interictal networks (reference group).  Subsequently, the proximity of nodes from a network to the interictal connectivity can be evaluated through the hyperbolic error ~\cite{barycenter_algo}.  Learning from the interictal networks, the most representative data in the dataset, alleviates the bias related to class-imbalanced training datasets. For each network in both the preictal and interictal group, the position of each node in the hyperbolic disk was examined and the node was assigned the value of the hyperbolic Gaussian probability density function of this node in the reference interictal group. The principle is illustrated in Fig.~\ref{fig:data_gauss}.  

It should be noted that hyperbolic space is a Riemannian manifold, rather than a vector space. As a result, the basic operations of vectors and matrices are either intractable or not explicitly defined in hyperbolic spaces. Nevertheless, for a point $z \in \mathbb{D}$, the tangent space at $z$, denoted by $T_z\mathbb{D}$, is an inner product space, which contains the tangent vector with all possible directions at $z$. We can transport points in $\mathbb{D}$ to the tangent space $T_z\mathbb{D}$, and apply most vector operators (average, gradients, \ldots) within this Euclidean subspace, and then project back the result to the disk. Transportation between $T_z\mathbb{D}$ and $\mathbb{D}$ can be achieved via the logarithmic and exponential maps, $\mathrm{Log}_{z}$ and $\mathrm{Exp}_{z}$, respectively,  defined in Appendix~\ref{App:TangLogSpace}.

To estimate the hyperbolic error we first projected each node $i$ from all reference (interictal) networks into a tangent space to estimate their barycenter $\hat{z}_i$ with the algorithm detailed in Appendix~\ref{App:BaryComp}. For each node $i^k$ from network $k$ in the second group, with coordinates $(x_i^k, y_i^k)$ in the hyperbolic disk $\mathbb{D}$, the score value was calculated using the following formula: 
\begin{align}
        \mathrm{Score}(i^k) & =  \mathrm{P}_i^{\mathrm{ref}}\left( {W_i^k}  \right) \nonumber\\
        & =  \frac{1}{\sqrt{\left( 2 \pi \right)^2\|V_i\|}} \mathrm{Exp} \left[ -\frac{1}{2} {{W_i^k}}^T V_i^{-1} {W_i^k} \right]
\end{align}  

where 
\begin{equation}
{W_i^k} = \begin{bmatrix}
                     \mathrm{Real} \left(\mathrm{Log}_{\hat{z}_i}\left(x_i^k + j y_i^k\right) \right)  \nonumber \\
                     \mathrm{Im} \left(\mathrm{Log}_{\hat{z}_i}\left(x_i^k + j y_i^k\right) \right)
                    \end{bmatrix}
\end{equation}
denotes the projected coordinates of node $i^k$ from network $k$ in the tangent space centered on $\hat{z}_i$ (the hyperbolic barycenter of nodes $i$ in the reference group). $\|V_i\|$ is the determinant of covariance matrix $V_i$ of the nodes $i$ from the reference embeddings projected in the tangent space centered on $\hat{z}_i$, and $j$ denotes the imaginary unit $j=\sqrt{-1}$.  It should be noted that the barycenter of the distribution of the set of points projected in this tangent space has coordinates $(0,0)$.

$\mathrm{P}_i^{\mathrm{ref}}\left( {W_i^k}  \right) $ is the hyperbolic Gaussian model of nodes $i$ in the reference interictal networks, estimated at the position of the $i^{\mathrm{th}}$ node of a network $k$. Once the scores of all the nodes have been calculated, the disk is discretized with a grid, and the errors of the nodes can be interpolated in the whole disk (one for each network). To perform the interpolation, the scores of the $n=8$ nearest nodes to each pixel of the grid are averaged, with the weights being inversely proportional to the distances between the pixel and the nodes' positions on the disk. This results in a Hyperbolic Score Disk (\rev{HSD}) for each network (see Fig.~\ref{fig:data_gauss}).
%===== FIGURE =====

\subsection{\label{sec:RoiDef}Discrimination in the Hyperbolic Score Disk}
To ascertain whether there were any significant differences between the hyperbolic score disks of the preictal and interictal groups, a Student's t-test was employed. Firstly, all HSDs from the two groups were obtained. For a statistical cross-validation of the preictal-interictal networks separation, the two groups can be a subsample of the total number of disks. With the two groups defined, all pixels in the HSD are vectorized, and compared with a t-test to obtain a vector of t-values (one for each pixel of the disk), which can then be back-projected onto the original hyperbolic disk. Permutation tests were employed to provide an accurate approximation of the $p_{value}$ for each pixel. These tests entailed randomly permuting individuals between the two groups and subsequently performing t-tests on the resulting groups to compare them with the original value. The region of interest (referred here to as the ROI) is the region of the disk where the $p_{values}$ are the smallest, indicating the region of the disk where the two groups exhibit the greatest statistical divergence. An example of ROI is illustrated in Fig.~\ref{fig:meth_pred}.

\subsection{Prospective forecasting}
%===== FIGURE =====
\begin{figure*}[ht]
\includegraphics[width=0.9\textwidth]{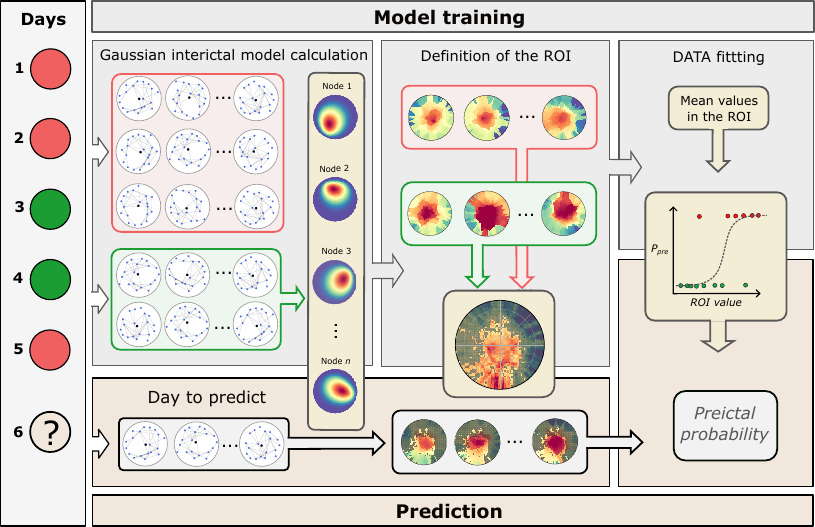} %Here is how to import EPS art
\caption{\label{fig:meth_pred} Forecasting method for one patient and one frequency band. In this example, the aim is to forecast the seizure likelihood of networks from day 6 using data from days 1 to 5. To achieve this, Gaussian models are associated to all nodes using interictal networks from days 1 to 5. The region of the HSD where the two groups exhibit the largest statistical difference (ROI) is then defined. The values within this ROI are utilized to train a logistic regression model, which predicts the category of each network of the following day(s).}
\end{figure*}

The model was trained using data from all previous days to predict the probability of a given network being in the preictal group (with a seizure within the next 24 hours). The training set was required to include at least one day from each group (preictal and interictal).  Gaussian models were constructed from the embedded interictal networks, and the hyperbolic score disk was calculated for all networks in the training set. The two groups of HSD (interictal and preictal networks) are then compared, and a region of interest (ROI) is identified within the hyperbolic disk, as previously described. For each embedded network from the learning set, the median value of the pixels within the ROI is calculated and used as an independent variable for the prediction. As the connectivity matrices were available in multiple frequency bands, these steps were repeated for each frequency band. Each disk (30 per day) was assigned to the class of the day (with or without seizures) and was associated with a vector containing the median values of the ROIs for each frequency band. The aforementioned data were subsequently utilized to fit a logistic regression, thereby enabling a new network to be assigned a probability, $p$, of belonging to a preictal day. For each new day, the probability of belonging to the preictal class was calculated by averaging the probabilities associated with the 30 disks of the day. The combinations of bands with the highest prediction score, calculated using the training set of each patient, were then utilized in a logistic regression for the prediction of successive days. It is important to note that the selected combination of bands may differ between patients. The methodology for a single frequency band is illustrated in Fig.~\ref{fig:meth_pred}.

\subsection{Evaluation of classification and forecasting  performances}
As connectivity patterns may be highly correlated between short epochs from the same daily recording, which could result in overestimated discrimination performances, we considered a leave-one-day-out cross-validation: for each patient, all 30 networks from a single day were assigned to the testing dataset, whereas the remaining data was used to train the algorithm. Once the model had been trained, it was tested on the data from the removed day, and its preictal probability were calculated. This process was repeated for all the days. 

\begin{table}[htb]
\caption{\label{tab:ResLeaveOneOut} Discrimination performances. Results indicate the mean values $\pm$ the standard deviation over the patients. They are presented for each frequency band considered independently, and the final row displays those obtained from the best combination of bands in each patient.}
\begin{ruledtabular}
\begin{tabular}{cccc}
Band     &  AUC Value                & F1 score                & Accuracy                \\
\hline
$\delta$        & $0.387 \pm 0.35$ & $0.33 \pm 0.34$ & $0.56 \pm 0.28$ \\
$\theta$        & $0.32 \pm 0.29$  & $0.29 \pm 0.32$ & $0.50 \pm 0.22$ \\
$\alpha$        & $0.623 \pm 0.26$ & $0.38 \pm 0.27$ & $0.62 \pm 0.09$ \\
$\beta$        & $0.46 \pm 0.40$  & $0.39 \pm 0.42$  & $0.53 \pm 0.33$ \\
Low $\gamma$        & $0.6675 \pm 0.31$ & $0.64 \pm 0.16$ & $0.75 \pm 0.16$ \\
High $\gamma$        & $0.277 \pm 0.35$  & $0.28 \pm 0.35$ & $0.55 \pm 0.21$ \\
Combined & $\mathbf{0.932 \pm 0.07}$ & $\mathbf{0.79 \pm 0.18}$ & $\mathbf{0.85 \pm 0.13}$
\end{tabular}
\end{ruledtabular}
\end{table}

%===== FIGURE =====
\begin{figure*}[ht!]
\centering
\subfloat{%
\includegraphics[width=0.45\textwidth]{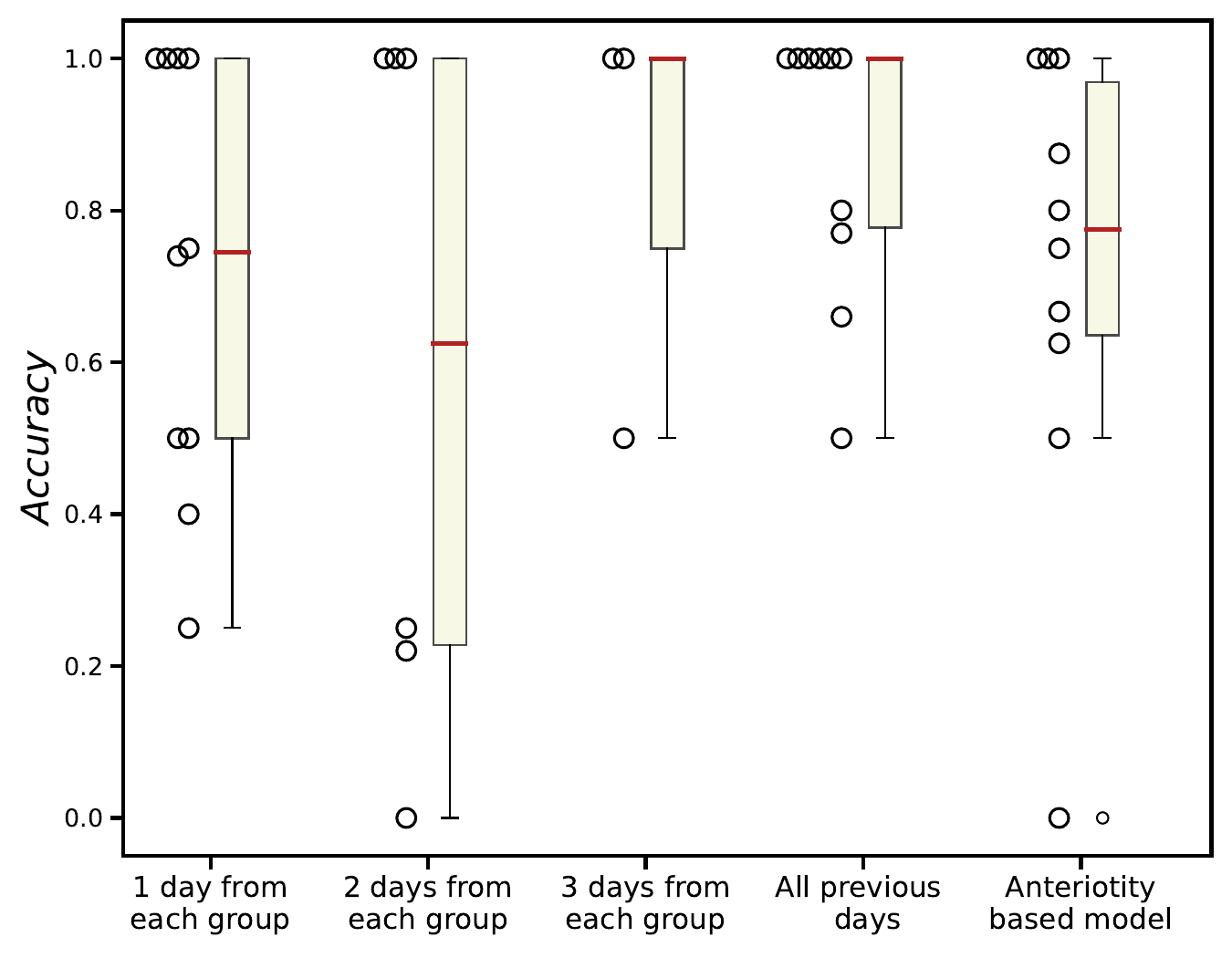}
}
\hspace{0.3 cm}
%\vspace{-10.5 cm}
\subfloat{%
\includegraphics[width=0.45\textwidth]{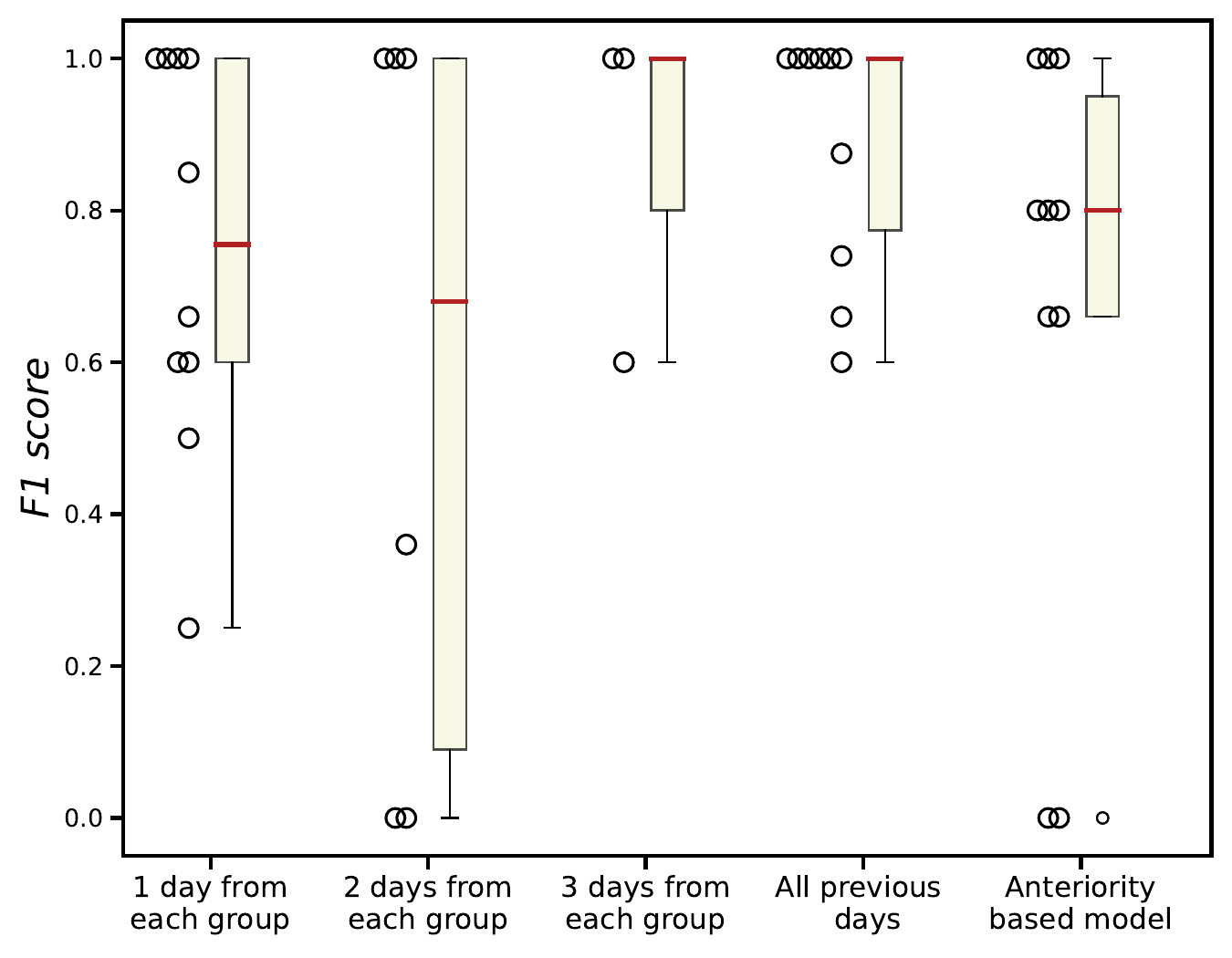}
}
\caption{\label{fig:AccF1} Accuracy and F1 score results based on the minimum number of days from each class in the training set. Data are presented in scatter plots where each small circle represents a patient, and the red line indicates the median. The term ``Anteriority-based model'' refers to the method that bases its predictions on the immediate previous day’s state.} 
\end{figure*} 

The discriminatory potential and forecasting performance of the model were evaluated based on several metrics, including accuracy, recall, precision, F1-score, and the area under the receiver operating characteristic curve (AUC). Accuracy measures the proportion of correct predictions made by the learning model. Recall or sensitivity quantifies the model's ability to identify all relevant positives. Precision, also known as positive predictive value, assesses the model's ability to distinguish true positives from false positives. To take into account the class imbalance of the dataset (i.e., the number of preictal days is lower compared to interictal epochs) we used here the F1-score, which is a weighted average of recall and precision. F1-score strikes a balance between minimizing false negatives and false positives, thereby ensuring a reliable assessment of the model's effectiveness. Finally, the AUC, which ranges from 0.5 for a random classification to 1 for a perfect classification, provides a comprehensive assessment of the model's performance.

To provide a comprehensive evaluation of our forecasting model we computed the Brier score (i.e., the mean squared error over every forecast; from 0 [perfect prediction] to 1 [worst prediction]) and the Brier skill score (BSS), which measures the improvement over a default prediction (given the limited number of recording sessions, the default probability of seizures was estimated as the proportion of the total preictal/interictal epochs for each patient), and take values from 0 (no improvement) to 1 (perfect forecasting), with negative values indicating worse performances than default predictions. 

\section{Results}
\subsection{Discrimination of preictal networks}
We first evaluated the ability of the proposed method to discriminate between interictal and preictal networks. Classification performances were assessed by AUC, the accuracy, and the F1-score and are presented in Tab.~\ref{tab:ResLeaveOneOut}. In general, the connectivity patterns associated with activities of separated frequency bands cannot distinguish between preictal and interictal days. Only networks estimated from low $\gamma$ oscillations (30-49~Hz) performed better than a random classifier. Nevertheless, classification performances considerably increased when the information from different frequency bands was combined for each patient:  All patients reached AUCs of $\geqslant 0.78$ (mean AUC $=0.93$, mean F1 score $=0.79$, and mean accuracy $=0.85$).  These findings suggest that combining connectivity across different frequency bands provides valuable information for distinguishing preictal from interictal epochs. \rev{Notably, the hyperbolic embedding method outperformed a standard machine learning model, specifically a support vector machine (SVM). The SVM model yielded an average AUC of $0.79$, an F1 score of $0.78$, and a mean classification accuracy of $0.77$, demonstrating the efficacy of the proposed hyperbolic network embedding approach in enhancing seizure risk prediction.}

\subsection{\label{sec:PseudoProsMeth}Forecasting of upcoming seizure(s)}
The forecasting method was first assessed by examining the number of days from each class in the training set. The results are presented in Fig.~\ref{fig:AccF1}, which plots the accuracy and F1 score against the minimum number of days from each class in the training set. When only one day ($n=1$) from each class was included in the training set, the average accuracy of prediction was $0.71$ [min. 0.25, max. 1] and the mean F1-score was $0.75$ [min. 0.25, max. 1]. However, these values increased considerably to $0.83$ [min. 0.5, max. 1] for the accuracy and $0.87$ [min. 0.61, max. 1] for the F1-score when the model was trained with three days from each class. When every day was predicted with a model trained with data from all previous days, performances reached the highest values of accuracy $0.87$ and F1-score $0.89$ (See Fig.~\ref{fig:AccF1}). The use of different reference networks (from previous days) in the alignement of the embedded data yielded similar performances.

%===== FIGURE =====
\begin{figure*}[ht]
\includegraphics[width=0.9\textwidth]{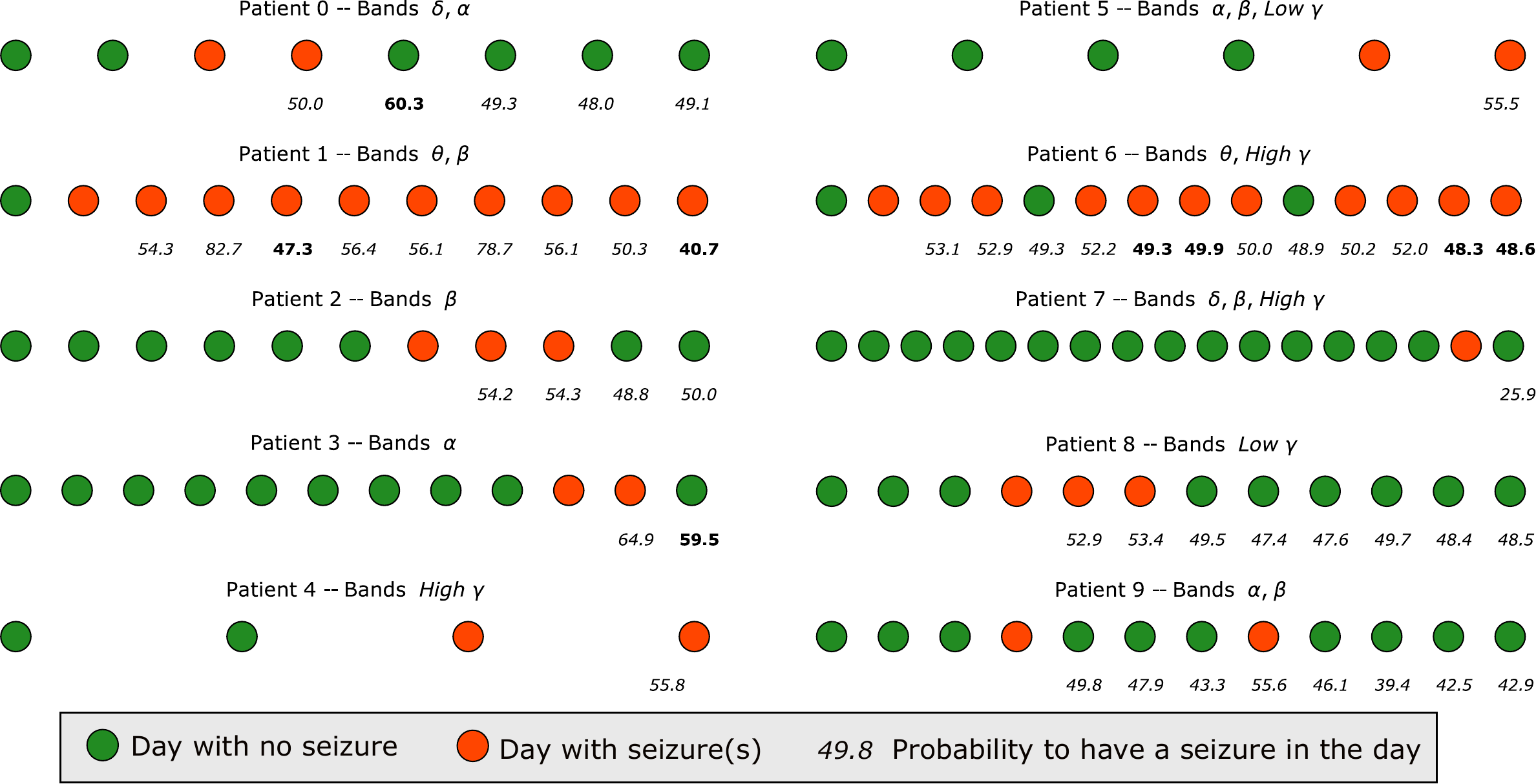} %Here is how to import EPS art
\caption{\label{fig:res_patients} Forecasting results for all patients using data from all previous days in the training model. Probabilities that give the wrong class to the day are highlighted in bold. }
\end{figure*}

Fig.~\ref{fig:res_patients} presents the forecasting results for all patients when data from all the previous days were used for training the model. Patient-specific predictions were performed with the most discriminant combination of networks in different frequency bands.  Of the 28 preictal days (days with seizures), 22 (78.5\%) were correctly predicted with a probability of $>50$\%.  False positive predictions were observed in only two of 23 (8.6\%) interictal periods (in two patients), and they exclusively occurred after two consecutive days with seizures.

The efficacy of our method was evaluated in comparison to a basic anteriority-based model, which is a classifier that bases its predictions on the previous day’s state. This simple model yielded an average accuracy value of $0.72$, but this value could be overestimated as accuracy does not take into account the class imbalance (i.e., the relatively reduced number of preictal days). In this case, the F1-score, which combines precision and recall, provides a better assessment of model performance. Here, the anteriority-based model yielded a moderate F1-score of $0.67$. Further, we noticed that this value was very close to that obtained with a non-informative (random), which yielded an averaged F1-score of $0.68$ (for each patient, this non-informative model’s F1-score can be estimated by $2p/(p+1)$, where $p$ is the rate of preictal epochs). 

Forecasting results were consistent with those of our previous study, in which we demonstrated that brain connectivity obtained from vigilance-controlled resting-state recordings could accurately predict the risk of upcoming seizures~\cite{24_Cousyn2023}.  The results, presented in Table.~\ref{tab:ResPred}, demonstrate the forecasting performance obtained for the distinct frequency bands when the prediction models for each day were trained with data from all preceding days. Combining information from different frequency bands yielded a good predictor with an average Brier score and BSS of $0.12$ [min. = 0.003, max. = 0.48] and $0.36$ [min. $=-0.43$, max. $=0.70$], respectively. \rev{In comparison, the SVM model achieved a mean Brier score of $0.13$}. It is worthy of notice that connectivity networks from the $\beta$ and low $\gamma$ band also provided good predictors, in agreement with previous findings~\cite{24_Cousyn2023}. \rev{These results demonstrate that our prediction model outperformed other approaches, including a standard machine learning model, in forecasting the risk of upcoming seizure(s).}  

\section{Discussion}
This study provided empirical evidence supporting the hypothesis that hyperbolic spaces are suitable for representing complex brain connectivity patterns. This makes it a promising framework for accurately distinguishing between interictal and preictal states. In contrast to standard seizure prediction methods that track changes in EEG dynamics by analyzing long-term recordings~\cite{15_Kuhlmann2018, beniczky2021machine}, our study confirmed that short and daily single resting-state recordings can reflect a pro-ictal state, at least at the daily level.

\begin{table*}%[b]
\caption{\label{tab:ResPred} Forecasting performances. The given results are the averaged values over the patients $\pm$ the standard deviation. The results are presented for each band considered independently and the final row displays the result for the best combination of bands for each patient. BS : Brier Score, BSS : Brier Skill Score. }
\begin{ruledtabular}
\begin{tabular}{lcccccc}
Band & \multicolumn{1}{l}{BS}    & \multicolumn{1}{l}{BSS}  & \multicolumn{1}{l}{F1 Score}   & \multicolumn{1}{l}{Accuracy} \\
\hline
$\delta$        	 & $0.10 \pm 0.12$    & $0.40 \pm 0.44$	 & $0.36 \pm 0.38$ 	&  $0.49 \pm 0.31$    \\
$\theta$          	 & $0.08 \pm 0.05$    & $0.52 \pm 0.41$ 	 & $0.53 \pm 0.35$ 	& $0.55 \pm 0.27$     \\
$\alpha$        	 & $0.08 \pm 0.04$    & $0.48 \pm 0.41$ 	 & $0.44 \pm 0.35$  	& $0.52 \pm 0.20$     \\
$\beta$        	 & $0.08 \pm 0.05$    & $0.44 \pm 0.41$ 	 & $0.67 \pm 0.34$ 	& $0.65 \pm 0.33$     \\
Low $\gamma$       & $0.08 \pm 0.06$    & $0.46 \pm 0.45$  & $0.70 \pm 0.31$ 	& $0.74 \pm 0.22$	     \\
High $\gamma$      & $0.08 \pm 0.06$    & $0.50 \pm 0.40$  & $0.46 \pm 0.35$ 	& $0.55 \pm 0.35$     \\
\textbf{Combined}  & $0.12 \pm 0.12$    & $0.36 \pm 0.35$  & $0.89 \pm 0.14$ 	& $0.87 \pm 0.17$
\end{tabular}
\end{ruledtabular}
\end{table*}

The classification performances demonstrated a high degree of discriminating power when the information from brain connectivity in different frequency bands was combined for each patient. The combination of networks from different frequency bands yielded the most accurate predictions (an accuracy of 87\% and F1-score of 89\%), but it is notable that connectivity networks from the beta and low gamma bands also demonstrated excellent forecasting results. For all patients, the predictive performance was enhanced by increasing the quantity of learning data. For each day, the optimal prediction was achieved when the models were recursively trained with networks from all preceding days. For these models, false positives were only detected in two patients, and only occurred immediately after consecutive preictal periods. This may indicate a pro-ictal condition, in which seizures are more likely to occur after a day of seizures. However, homeostatic control mechanisms could prevent their emergence. 

\rev{In contrast to existing models that depend on long-term EEG recordings to monitor preictal dynamics, our findings demonstrate that brief resting-state EEG segments can yield clinically relevant information regarding the preictal state on a day-to-day basis. It is worth noting that a direct comparison to traditional seizure prediction methods is not entirely feasible due to methodological differences, particularly in the duration of recordings utilized. However, compared to alternative models such as the SVM classifier~\cite{24_Cousyn2023} or a simple anteriority-based model, our approach exhibits superior performance. Additionally, most EEG-based methods for seizure prediction provide forecasts within minutes to, at most, an hour before a seizure event~\cite{Usman2019, Rasheed2021}. In contrast, our method provides a single daily forecast, offering an extended time window for intervention or monitoring throughout the day.}

It should be noted that other non-Euclidean embeddings (e.g. spherical or elliptic), and that alternative dimensional reduction methods (e.g., Isomaps, Locally Linear embeddings~\cite{18_vonLuxburg2007}) can also be used. Similarly, representations in different hyperbolic spaces, such as Lorentz-Klein, $d-$dimensional Poincaré's ball, or the half-space model, can be used as alternatives to Poincaré's disk. Although other hyperbolic mappings (e.g., Mercator~\cite{GarciaMercator1019} HyperMap~\cite{papadopoulos2014network} or Hydra~\cite{keller2020hydra} among others) are alternatives to projecting brain networks into the 2D Poincaré disk, the coalescent embedding method used here encompasses various benefits. These include a short computational time and the absence of stochastic minimization procedures, thus preventing errors introduced by local minima.  \rev{Similarly, while alternative methods based on the Minimum Spanning Tree (MST) can be used for filtering connectivity matrices~\cite{luppi2024systematic},  the MST~\cite{stam2014trees} and MST+ECO~\cite{8_DeVicoFallani2017} approaches exhibit poor performance in forecasting preictal states, yielding an average accuracy $\leq 0.56$ and a F1-score $\leq 0.4$. These results highlight the advantages of our chosen filtering method.}

\revb{The application of hyperbolic geometry for embedding brain networks offers novel insights into the structural organization of functional connectivity. Hyperbolic space is inherently suited to represent hierarchical and complex network architectures, characteristics commonly observed in biological systems such as the human brain. In our model, hyperbolic distances between nodes encode both the strength and the hierarchical nature of inter-regional connections, enabling the detection of network alterations that may be overlooked by Euclidean or other geometric representations. Our findings indicate that this geometric framework yields a precise characterization of connectivity changes, including those associated with epileptogenic processes.}

\revb{While other non-Euclidean methods, such as Riemannian embeddings of covariance matrices, have been extensively employed in EEG analysis (particularly within brain-computer interface applications), the use of hyperbolic geometry to model functional connectivity, especially in the context of epilepsy, remains relatively underexplored. This work represents, to our knowledge, the first application of such embeddings for seizure prediction, opening promising perspectives for future studies.}

%Despite the good classification and prediction results, our study has some limitations. First, the vigilance-controlled resting-state iEEG recordings were obtained from hospitalized patients who were candidates for presurgical evaluation. Although no trivial correlations were found between the occurrence of seizures and the controlled medication tapering~\cite{24_Cousyn2023}, the results cannot be directly generalized to out-of-hospital real-life conditions. Similarly, given the heterogeneity of electrode location among the patients, we cannot generalize a single unique prediction model. While a larger cohort of patients is necessary to map out the limits of the proposed approach, the results do suggest that the method has the potential to be a promising tool for the analysis of brain networks.

\revb{Despite the promising classification and prediction performance demonstrated by our model, several limitations should be acknowledged. First, due to the heterogeneity in electrode implantation sites across patients, the development of a generalized, patient-independent prediction model remains infeasible at this stage. Second, the vigilance-controlled resting-state iEEG recordings were obtained from hospitalized individuals undergoing presurgical evaluation. Although no significant associations between seizure occurrence and medication tapering were identified in the original dataset~\cite{24_Cousyn2023}, the potential influence of time-dependent factors and treatment adjustments cannot be entirely excluded—an inherent limitation in such clinical environments. Consequently, the generalizability of our approach to out-of-hospital or real-world scenarios is limited, as these contexts involve greater long-term variability and more heterogeneous clinical profiles.}

\revb{The inclusion of larger datasets with vigilance-controlled recordings at different times of day could facilitate evaluation of diurnal variations in seizure risk, thereby enhancing clinical relevance. Furthermore, integrating additional physiological signals—such as heart rate variability, spontaneous baroreflex sensitivity, and electrodermal activity—may further improve predictive performance.}

\revb{To summarize, our findings indicate that hyperbolic embedding of brain networks captures discriminative features between interictal and preictal states, offering a promising framework for accurate daily seizure risk forecasting.}

\begin{acknowledgments}
M.G. acknowledges doctoral support from the Ecole Normale Supérieure de Lyon, France.
\end{acknowledgments}

\appendix
\section{\label{App:HypGausDist}Hyperbolic Gaussian distribution}
Let us consider an ensemble of points in the Poincaré's disk. As in the case of the Euclidean space, the distribution of the points can be characterized by a barycenter and a covariance matrix. This information is sufficient to approximate the distribution by a two-dimensional Gaussian distribution. 

\subsection{\label{App:TangLogSpace}Hyperbolic logarithmic and exponential maps}
The logarithmic map $\mathrm{Log}_u(v)$ provides a way to project a point $v \in \mathbb{D}$ (in complex coordinates) into the tangent space $T_u\mathbb{D}$ at the point $u$,  as follows :
\begin{equation}
    \mathrm{Log}_u(v) = (1-|u|^2)\mathrm{atanh}\left( \left| \frac{v-u}{1-\bar{u}v} \right|\right)e^{j\theta}
\end{equation}
where $\theta = \mathrm{arg}\left( \frac{v-u}{1-\bar{u}v} \right)$, and $\mathrm{atanh}(\cdot)$ denotes to the inverse of the hyperbolic tangent function.

The exponential map $\mathrm{Exp}_u(v)$ allows to do the inverse transformation, i.e. projecting a point $v \in T_u\mathbb{D}$ from the tangent space at $u$ into the disk $\mathbb{D}$ in the following way: 
\begin{equation}
    \mathrm{Exp}_u(v) = \frac{\left( u + e^{i\theta} \right) e^{2\|v\|} + \left( u - e^{j\theta} \right) }{\left( 1 + \bar{u}e^{j\theta} \right) e^{2\|v\|} + \left( 1 - \bar{u}e^{j\theta} \right) }
\end{equation}
where $\theta = \mathrm{arg}(v)$ and $\|v\| = \frac{|v|}{1-|u|^2}$. \\

\subsection{\label{App:BaryComp} Hyperbolic barycenter computation}
To estimate the barycenter of a set of points in the Poincaré disk, we used the algorithm presented in Ref.~\cite{barycenter_algo}. The algorithm converges to the barycenter of the points by means of recursive projections in the tangent space. It iterates between the Poincaré disk and the tangent space until convergence is achieved. The algorithm is presented in Algorithm.~\ref{alg:bary_comp}.

\subsection{Estimation of covariance matrix in the hyperbolic disk}
To estimate the covariance matrix of a set of points $z_i$ in the hyperbolic disk, we first calculated their barycenter $u$ with the algorithm described above. The points distribution is then projected into the tangent space centered on the estimated barycenter $\hat{z}$ of the points on $T_u\mathbb{D}$, using the logarithmic map $\mathrm{Log}_{\hat{u}}(\cdot)$. In this Euclidean subspace, the barycenter is the point with coordinates $(0, 0)$ and the covariance matrix can finally be calculated as follows: $V_{\hat{z}} = \mathrm{Covariance} \left( \left[ \mathrm{Log}_{\hat{z}}(z_i) \right]^T \right)$

\SetKwComment{Comment}{/* }{ */}
\RestyleAlgo{ruled}
\begin{algorithm}[ht]
\caption{Barycentre computation}\label{alg:bary_comp}
\KwData{\\ $\quad  Z = \{ z_i, 1 \leq i \leq n\} $ : complex coordinates of the points in the Poincaré's disk, \\ 
\quad $z_{init}$ : Barycentre initialisation (complex coordinates), \\ 
$\quad \tau_b$ : step size (strictly positive), 
\\ $\quad d_{\mathrm{thres}}$ : threshold of convergence for $d$.}
\KwResult{\\ $\quad \hat{z}$ : numerical approximation of the barycenter (in complex coordinates) \\}
\vspace{0.2 cm}
$\hat{z} \gets z_{init}$ \;
$d \gets 1\mathrm{e}6$ \;
$N \gets n$ \;
\While{$d \geq d_{thres}$}{
$\mu \gets \frac{2}{n} \sum_{i=1}^n \mathrm{Log}_{\hat{z}}\left( z_i \right)$ \;
$\hat{z} \gets \mathrm{Exp}_{\hat{z}}\left( \tau_b \mu \right)$ \;
$d \gets \sqrt{\frac{|\mu|^2}{\left(1 - |\hat{z}|^2 \right)^2}}$ \;
}
\end{algorithm}

\bibliography{paperHyperbolicSeizurePrediction_V14062024}% Produces the bibliography via BibTeX.
\end{document}